\documentclass[aps,prl,twocolumn,footinbib,superscriptaddress]{revtex4-2}

\pdfoutput=1

\usepackage[utf8]{inputenc} 
\usepackage{amsmath}
\usepackage{amssymb}
\usepackage{xcolor}
\usepackage{graphicx}
\usepackage{subfigure}
\usepackage{dcolumn}
\usepackage{bm}
\usepackage{microtype}
\usepackage{threeparttable}
\usepackage{mathtools}
\usepackage[colorlinks=true]{hyperref}
\usepackage{physics}
\usepackage{comment}
\usepackage{soul} 
\usepackage{bbold} 
\usepackage{ragged2e}
\usepackage{graphicx,overpic}
\usepackage[most]{tcolorbox} 
\usepackage{todonotes} 
\usepackage{framed}

\tcbset{highlightstyle/.style={
  colback=yellow!20,
  colframe=yellow!80!black,
  boxrule=0pt,
  arc=2pt,
  outer arc=2pt,
  boxsep=2pt,
  left=2pt,
  right=2pt,
  top=2pt,
  bottom=2pt
}}

\newcommand{\del}{\partial}

\newcommand{\be}{\begin{eqnarray}}
\newcommand{\ee}{\end{eqnarray}}

\begin{document}

\title{De Sitter holographic complexity from  Krylov complexity in DSSYK}

\author{Michal P. Heller}
\email{Michal.P.Heller@UGent.be}
\affiliation{Department of Physics and Astronomy, Ghent University, 9000 Ghent, Belgium}
\affiliation{Institute of Theoretical Physics and Mark Kac Center for Complex Systems Research,
Jagiellonian University, 30-348 Cracow, Poland}

\author{Fabio Ori}
\email{Fabio.Ori@UGent.be}
\affiliation{Department of Physics and Astronomy, Ghent University, 9000 Ghent, Belgium}

\author{Jacopo Papalini}
\email{Jacopo.Papalini@UGent.be}
\affiliation{Department of Physics and Astronomy, Ghent University, 9000 Ghent, Belgium}

\author{Tim Schuhmann}
\email{Tim.Schuhmann@UGent.be}
\affiliation{Department of Physics and Astronomy, Ghent University, 9000 Ghent, Belgium}

\author{Meng-Ting Wang}
\email{MengTing.Wang@UGent.be}
\affiliation{Department of Physics and Astronomy, Ghent University, 9000 Ghent, Belgium}

\begin{abstract}
We utilize the recent connection between the high energy limit of the double-scaled SYK model and two-dimensional de Sitter solutions of sine dilaton gravity to identify the length of a family of geodesics spanned between future and past infinities with Krylov spread complexity. This constitutes an explicit top-down microscopic realization of holographic complexity in a cosmological spacetime. Our identification is different from the existing holographic complexity proposals for de Sitter geometries which are anchored either on horizons as holographic screens or on timelike observers. This leads us to introduce and investigate a new cosmological holographic complexity proposal in any dimension. It is based on extremal timelike volumes anchored at the asymptotic past and future and at large values of the anchoring boundary coordinate grows linearly with growth rate proportional to the product of de Sitter entropy and temperature.
\end{abstract}

\maketitle

\noindent \textbf{\emph{Introduction.--}} Studying the connection between information processing and dynamical spacetime dates back to pioneering works on black hole thermodynamics in the 1970s~\cite{Bekenstein:1973ur,Hawking:1976de} and has witnessed a true renaissance over the course of the past two decades, see~\cite{Rangamani:2016dms,Almheiri:2020cfm,Chen:2021lnq,Baiguera:2025dkc} for reviews. The contemporary explosion of results in this area has a twofold origin. On one hand, it is formulating holography --- a precise microscopic correspondences between gravitational theories and quantum mechanical systems with large number of strongly-interacting constituents~\cite{Maldacena:1997re,Witten:1998qj,Gubser:1998bc}. On the other, within holography, it is establishing  a fully quantitative relation between microscopic entanglement entropy and areas of certain extremal surfaces~\cite{Ryu:2006bv,Hubeny:2007xt,Casini:2011kv,Lewkowycz:2013nqa,Dong:2016hjy}.

This shift of the research perspective to view black holes as quantum information processing objects has led to a realization about a possibility of a novel connection  between spacetime volumes and some notion of complexity of black hole states~\cite{Susskind:2014rva,Stanford:2014jda,Brown:2015bva,Couch:2016exn,Belin:2021bga}. The emerging field of holographic complexity was recently put on a much firmer footing by identifying an explicit notion of complexity, the Krylov spread complexity~\cite{Balasubramanian:2022tpr,Nandy:2024evd,Rabinovici:2025otw}, that acquires a precise interpretation~\cite{Lin:2022rbf,Rabinovici:2023yex,Heller:2024ldz} in terms of a volume in the simplest incarnation of a holographic correspondence for two-dimensional gravity. This is important as it builds a concrete bridge between ideas pursued in the context of operator growth and the physics of the black hole interior. 

While at a superficial level the information processing picture of gravity seems to apply to any theory of gravity~\cite{tHooft:1993dmi,Susskind:1994vu}, so far the microscopic understanding pertained to holography in asymptotically Anti-de Sitter (AdS) spacetime. 
This limitation applies in particular to positively curved spacetimes, which model the accelerated expansion of our universe in the future. Nevertheless, over the course of the past several years, efforts were made to generalize both holographic entanglement entropy~\cite{Shaghoulian:2022fop,Doi:2022iyj} and holographic complexity to de Sitter (dS) universes~\cite{Reynolds:2017lwq,Susskind:2021esx,Chapman:2021eyy,Chapman:2022mqd,Jorstad:2022mls,Anegawa:2023wrk,Aguilar-Gutierrez:2023zqm,Anegawa:2023dad,Baiguera:2023tpt,Auzzi:2023qbm,Baiguera:2024xju,Mohan:2025aiw}. Our work addresses the latter part of this broader research program.

The core idea of the present article is to build on the precise correspondence between the double-scaled SYK model (DSSYK) and two-dimensional sine dilaton gravity \cite{Blommaert:2024ydx}. 
This theory of gravity has limits that describe both two-dimensional negatively, as well as positively curved spacetime, where the latter is the focus of this Letter.
We will utilize the recently established connection between the Krylov spread complexity with geodesic length in the effectively negatively curved regime~\cite{Rabinovici:2023yex,Heller:2024ldz} and report how to map it to a positively curved original geometry. 
In this way we will establish that the Krylov spread complexity gives rise to a holographic notion of complexity in a two-dimensional cosmology. We will extrapolate from this result and propose a de Sitter-specific holographic complexity proposal that in two dimensions reduces to this top-down notion and in higher dimensions probes de Sitter geometries in analogous ways. 
\\

\noindent \textbf{\emph{Complexity dictionary for sine dilaton gravity dual to the DSSYK model.--}}   In order to establish our main model of study, let us introduce the essentials of the duality between disk level sine dilaton gravity and the double-scaled SYK model proposed in \cite{Blommaert:2024ydx} and subsequently elaborated on in \cite{Blommaert:2024whf,Blommaert:2025avl,Bossi:2024ffa,Blommaert:2025eps}. Sine dilaton gravity is a theory of dilaton gravity in two dimensions whose path integral
 \begin{equation}
 \label{eq:PI_sine_dilaton}
    \int \mathcal{D} g \mathcal{D}\Phi\,\exp\bigg[ \frac{1}{4\abs{\log q}}\int \mathrm{d}^2 x \sqrt{g}\bigg(\Phi R+2\sin(\Phi)\bigg)\bigg]
\end{equation}
includes a sine potential for the dilaton \footnote{Here, we have omitted writing out the GHY boundary term and the holographic renormalization counterterm.}. The parameter $\hbar=2\abs{\log q}$ with $q\in[0,1]$ governs the strength of quantum effects and $q\to1$ is the classical limit. The metric solution $ds^2$ on the classical saddle of \eqref{eq:PI_sine_dilaton} has curvature $R=-2\cos(\Phi)$. One key insight of \cite{Blommaert:2024ydx}, which will be important in this Letter, is that a special role is played by the Weyl-rescaled effective metric
\begin{align}\label{weyl}
    ds^2_{\text{eff}}=e^{-i\Phi}ds^2\,.
\end{align}
This metric has uniform negative curvature $R_{\text{eff}}=-2$ and describes an AdS$_2$ black hole with temperature $\beta_{\text{BH}}(\theta)=2\pi/\sin(\theta)$ \cite{Blommaert:2024ydx}.
The parameter $\theta$ is the value of the dilaton at the black hole horizon and characterizes the temperature of the dual DSSYK model. In order to establish the duality with the DSSYK model, sine dilaton gravity is canonically quantized in the effective geometry geodesic length $L_{\text{eff}}$ and its canonical conjugate $P_{\text{eff}}$. It is shown in \cite{Blommaert:2024ydx} that the gravitational Hamiltonian $\hat{H}_\text{grav}(\hat{L}_{\text{eff}},\hat{P}_{\text{eff}})$ that arises from quantizing the ADM energy coincides with the transfer matrix of the chord description \footnote{The auxiliary chord Hilbert space $\ket{n},\, n\in\mathbb{N}_0$ together with its transfer matrix $\hat{T}_\text{DSSYK}$, can be used to evaluate ensemble averaged moments of the Hamiltonian of the SYK model
\begin{equation}
\label{eq:H_syk}
H_{\mathrm{SYK}} = i^{p/2} \underset{{1 \leq i_1 < \cdots < i_p \leq N}}{\mbox{{\Large $\Sigma$}}} J_{i_1 \cdots i_p} \psi_{i_1} \cdots \psi_{i_p} \,,
\end{equation}
in the double-scaling limit where $p\to\infty$ and $N\to\infty$ while the ratio $\abs{\log q}=p^2/N$ is kept fixed. Here $J_{i_1 \cdots i_p}$ are Gaussian random couplings and $\psi_{i_1} \cdots \psi_{i_p}$ describes $p$-local Majorana fermion interaction.} \cite{Berkooz:2018jqr,Berkooz:2018qkz,Berkooz:2024lgq} of the DSSYK model \cite{Sachdev:1992fk,kitaevvideo}
\begin{align}
\label{eq:dictionary_H}
    \hat{H}_\text{grav}=\hat{T}_{\text{DSSYK}}\,.
\end{align}
The holographic dictionary
\begin{align}
\label{eq:dictionary_L}
    \hat{L}_\text{eff}=2\abs{\log q}\, \hat{n}    
\end{align}
relates fixed effective length states in sine dilaton gravity with the chord number states in the DSSYK model $\ket{L_{\text{eff}}}=\ket{n}$. Crucially, the discreteness and positivity of the length states $\ket{L_{\text{eff}}}$ arises naturally from gauging the shift symmetry $P_\text{eff}\to P_\text{eff}+2\pi$ of $\hat{H}_\text{grav}$, and also thereby recovers the true DSSYK temperature \cite{Blommaert:2024whf},
\begin{equation}\label{dssyk_temp}
\beta(\theta)=(2\pi-4\theta)/\sin(\theta).
\end{equation}
Together, \eqref{eq:dictionary_H} and \eqref{eq:dictionary_L} establish a dual description of observables in sine dilaton gravity as observables in the chord formulation of the DSSYK model.\\

For sine dilaton gravity, in \cite{Heller:2024ldz}, three of us showed that the expectation value of \eqref{eq:dictionary_L} in the time-evolved finite-temperature Hartle-Hawking state implies a particular complexity dictionary. The bulk gravitational interpretation of this expectation value is the quantum generalization 
of the spacelike wormhole length in the AdS$_2$ BH effective geometry. Because geodesic length in two dimensions coincides with the volume of the extremal codimension-1 boundary-anchored slice, introduced in \cite{Susskind:2014rva,Stanford:2014jda} as a bulk notion of complexity $\mathcal{C}$, the bulk interpretation of this expectation value is volume complexity
\begin{align}\label{CVsine_dilaton}
   \langle\hat{L}_\text{eff} \rangle=\mathcal{C}_\text{eff}(t)_\theta\,.
\end{align}
We will also equivalently refer to this notion of bulk volume complexity as the wormhole volume or Einstein-Rosen-Bridge (ERB) volume. The subscript $\mathcal{C}_\text{eff}$ is to remind the reader that we work with the effective geometry and hence interpret this as holographic complexity in the effective geometry. From a boundary viewpoint, on the other hand, the  expectation value \eqref{CVsine_dilaton} can be rewritten as a Krylov spread complexity -- introduced in \cite{Balasubramanian:2022tpr} by quantifying the spread of a state over its minimal basis of evolution or Krylov basis -- in the boundary DSSYK theory. For details of this derivation we refer to \cite{Heller:2024ldz} and the Supplemental Material. To summarize, \eqref{eq:dictionary_L} implies that volume complexity in sine dilaton gravity on the quantum level equals DSSYK Krylov spread complexity,
\begin{align}
\label{eq:c_duality_general}
    2 \abs{\log q}\,C_K(t)_\theta=\mathcal{C}_\text{eff}(t)_\theta\,.
\end{align}
For later reference, the classical limit $q\to1$ of this expression takes the simple form \footnote{Notice the well-known feature in sine dilaton gravity and the DSSYK model that, although the theory is at temperature $\beta(\theta)$ given in \eqref{dssyk_temp}, Lorentzian correlators, and hence also the length, scale in $t/\beta_{\text{fake}}(\theta)=t/(2\pi/\sin\theta)$, as discussed in \cite{Blommaert:2024ydx}, with properties  $\beta_{\text{fake}}(\theta)\geq2\pi$ and $\beta_{\text{fake}}(\theta)=\beta_{\text{fake}}(\pi-\theta)$.}
\begin{align}
\label{eq:c_duality_classicallimit}
    &\lim_{q\to1} 2\left|\log q\right|C_K(t)_{{\theta}}=\lim_{q\to1} \mathcal{C}_{\text{eff}}(t)_{{\theta}}\nonumber\\&=2 \log \left[\cosh \left(t\,\sin\theta /2 \right)\right]-2\log [\sin \theta]\,.
\end{align}
Before moving on, we highlight the key detail that will enable us to learn more about dS from this result. For now we have focused our attention on the \textit{effective} AdS geometry of sine dilaton gravity. As we will explain now, at least in certain limits, one can alternatively also work directly with the \emph{original} geometry and inquire about the geometric manifestation of microscopic complexity there.\\ 

\noindent \textbf{\emph{The dS limit of sine dilaton gravity.--}} 
In order to uncover a geometric manifestation of microscopic Krylov spread complexity in a cosmological spacetime in two-dimensional dilaton gravity, let us now introduce how such a spacetime is embedded in sine dilaton gravity. This has recently been discussed in \cite{Blommaert:2024whf,Blommaert:2025rgw,Okuyama:2025hsd}. We will revisit key arguments here and also present new understanding of the role of the Weyl rescaling that translates the original geometry into the effective geometry.

At the classical level, a simple way to appreciate the embedding of certain spacetimes into sine dilaton gravity is to examine limits of the classical metric solutions (the \textit{original} geometry) of \eqref{eq:PI_sine_dilaton}. The Lorentzian version of the classical sine dilaton gravity metric is given by \cite{Blommaert:2024ydx}
\begin{align}\label{sine}
    ds^2 = -\big( 2\cos\theta-2\cos\Phi \big)\,dt^2 
    + \frac{d\Phi^2}{ 2\cos\theta-2\cos\Phi }\, ,
\end{align}
where we have chosen the gauge in which the dilaton $\Phi=r$ parametrizes the radial coordinate \cite{Witten:2020ert}. It is well known that in the regime $\Phi \ll 1$, $\theta \ll 1$ the sine dilaton gravity solution \eqref{sine} reduces to the JT gravity solution \cite{Blommaert:2023wad,Blommaert:2024ydx}. This is the sine dilaton perspective on the earlier results of \cite{Berkooz:2018jqr,Lin:2022rbf} that JT gravity arises by zooming in on the lower edge $\theta\ll1$ of the continuous and bounded spectrum of the DSSYK model parametrized by $\theta\in[0,\pi]$. In this limit, the Weyl rescaling factor in \eqref{weyl} becomes trivial: the effective geometry coincides with the original sine dilaton geometry.

\begin{figure}
    \centering
    \includegraphics[width=\linewidth]{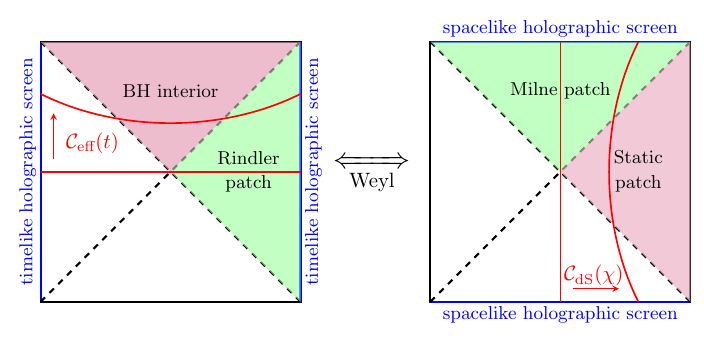}
    \caption{Parts of Penrose diagrams of maximally extended AdS$_2$ (\textbf{Left}) and dS$_2$ (\textbf{Right}) relevant to our setup. 
    Weyl rescaling the metric relates the AdS BH interior to the dS Static patch (red), the AdS Rindler patch to the dS Milne patch (green), the AdS BH horizon to the dS cosmological horizon (dashed), and the AdS asymptotic spatial boundaries to dS future and past boundaries (blue). As depicted in red, AdS extremal codimension-1 spacelike slices evolving in boundary time $t$ are mapped to timelike extremal codimension-1 slices anchored at past and future infinity that evolve with the spacelike boundary coordinate $\chi$ in dS. Full Penrose diagrams of the maximal extensions are discussed in \cite{Harlow:2018tqv} for AdS$_2$ and in \cite{Maldacena:2019cbz} for dS$_2$.}
    \label{fig:Ads_vs_dS}
\end{figure}

Let us now move on to the dS limit. 
In \cite{Blommaert:2024whf}, a different regime of sine dilaton gravity was studied, where the expansion is performed around the \emph{upper edge} ($\theta\simeq\pi$) of the DSSYK spectrum. One considers the limit 
\begin{align}
    &\Phi \;=\; \pi - \Phi_{\mathrm{dS}}, \qquad \Phi_{\mathrm{dS}}\ll 1,\\
    & \theta \;=\; \pi - \theta_{\mathrm{dS}}, \qquad \theta_{\mathrm{dS}}\ll 1 . \label{limit}
\end{align}
We will now omit the dS label. Expanding the original sine dilaton metric \eqref{sine} in this regime, one finds  
\begin{align}\label{dS1}
    ds^2 \;\overset{\Phi,\theta \simeq \pi}{\Longrightarrow}\;  -\big(\theta^2-\Phi^2\big)\,dt^2 
    \;+\; \frac{1}{ \theta^2-\Phi^2 }\,d\Phi^2 .
\end{align}
This is precisely a $-$AdS$_2$ metric, where the overall sign gives this geometry \emph{positive curvature} $R=2$. Equivalently, the metric \eqref{dS1} describes different patches of dS$_2$. For $\Phi>\theta$, the metric covers a \emph{Milne patch} of dS$_2$
\begin{align}
\label{eq:milne}
    ds^2_\text{Milne}=-d\tau^2+\sinh^2(\tau)d\chi^2\,,
\end{align}
while for $\Phi<\theta$ it covers a \emph{static patch} of dS$_2$ \footnote{What we show on the right side in Fig. \ref{fig:Ads_vs_dS} is accomplished from \eqref{dS1} with $\Phi\in(0,\infty)$ and $t\in(-\infty,\infty)$, by setting $\Phi = \theta \cosh(\tau)$ and $t = \chi / \theta$ for the Milne patch with $\tau\in(0,\infty)$, $\chi\in(-\infty,\infty)$, and $\Phi = \theta \cos(\phi)$ and $t = t_s / \theta$ for the static patch with $\phi\in(0,\pi/2)$, $t_s\in(-\infty,+\infty)$.}
\begin{align}
\label{eq:static}
    ds^2_{\text{static}}=-\sin^2(\phi)dt_{s}^{2}+d\phi^2\,,
\end{align}
based on the nomenclature of \cite{Maldacena:2019cbz}. $\Phi=\theta$ marks a \emph{cosmological horizon}, separating the static patch from the expanding Milne universe. See the right side of Fig.~\ref{fig:Ads_vs_dS}. The classical emergence of a dS$_2$ geometry in the described limit can also be seen directly at the level of the sine dilaton action, which reduces to dS JT gravity \footnote{Up to a topological term that can be absorbed into the Euler characteristic. Integrating out the dilaton indeed enforces $R=2$.}.

Similar to the JT case, this equivalence can furthermore be uplifted to the full quantum level \cite{Blommaert:2024whf}, since the sine dilaton/DSSYK partition function at $\theta \simeq \pi$ reproduces the wavefunction of the universe in dS JT gravity \cite{Maldacena:2019cbz,Cotler:2019nbi}.
In addition, \cite{Okuyama:2025hsd} proved that the full ETH matrix model for DSSYK \cite{Jafferis:2022wez}, which was proposed in \cite{Blommaert:2025avl} to be dual to sine dilaton gravity on higher topologies, reduces precisely to the matrix model of dS JT gravity \cite{Cotler:2024xzz,Maldacena:2019cbz}. This provides another nontrivial check of the identification of this regime of sine dilaton gravity with dS gravity. We provide some additional details on the quantum aspects of dS derived from this duality in the Supplemental Material.

A natural next step is now to examine the role of the Weyl rescaling \eqref{weyl} that maps the original geometry to the effective one.  Remarkably, in the $\Phi \simeq \pi$ regime, the rescaling simply contributes an additional minus sign \footnote{Notice that for general values $\theta$, away from the AdS and dS JT limits, the Weyl rescaling is a nontrivial function of the radial coordinate and the holographic contour where DSSYK lives in the effective geometry is mapped into a nontrivial complex contour in the original sine dilaton geometry \cite{Blommaert:2024ydx}. The Weyl factor is instead much simpler in the regimes considered in this Letter.}.
Thinking about the previously introduced patch structure of dS$_2$, the Weyl rescaling then acts by mapping the cosmological horizon into an effective black hole horizon, the static patch of the original dS$_2$ geometry into the BH interior of the effective AdS$_2$, and the
Milne patch of dS$_2$ \eqref{eq:milne} into the Rindler patch of AdS$_2$:
\begin{align}
\label{eq:milne_to_rindler}
    (-1)\,ds^2_\text{Milne}=-\sinh^2(\rho)dt^2+d\rho^2=ds^2_\text{Rindler}\,,
\end{align}
where we renamed $\tau _{\text{dS}}\leftrightarrow\rho_{\text{AdS}}$ and $\chi_{\text{dS}}\leftrightarrow t_{\text{AdS}}$ to match the standard AdS notation. The mapping is illustrated in Fig.~\ref{fig:Ads_vs_dS}, where corresponding patches are shown in the same color, and can intuitively be understood as a 90-degree rotation of the effective AdS diagram with respect to the dS diagram \footnote{The mappings are defined while keeping the Lorentzian signature $(-,+)$ fixed, which in turn determines the corresponding identification of a coordinate as spatial or temporal.}. Eq.~\eqref{eq:milne_to_rindler} emphasizes that the two asymptotic boundaries of the effective AdS$_2$ geometry, located at spatial infinity $\abs{\rho} = \infty$ in the two Rindler wedges, map to the past and future boundaries at $\abs{\tau} = \infty$ in dS$_2$. This precisely reproduces the boundary locations considered so far in the context of dS JT gravity \cite{Cotler:2019nbi,Maldacena:2019cbz,Cotler:2024xzz}. Furthermore, Eq.~\eqref{eq:milne_to_rindler} tells us to reinterpret the boundary time in the effective AdS geometry (which coincides with time $t$ in the DSSYK model, as discussed in \cite{Blommaert:2024ydx} and the previous sections) as a space coordinate at future and past infinity in the original geometry:
\begin{align}
\label{eq:t=chi}    t=t_{\text{Rindler,\,eff.\,AdS$_2$}}\leftrightarrow\chi_{\text{Milne,\,dS$_2$}}\,.
\end{align}

The map \eqref{eq:t=chi} can be applied to the aforementioned volume complexity $\mathcal{C}_{\text{eff}}$, defined by spacelike geodesics anchored on the two asymptotic boundaries of the effective AdS geometry at given time $t$. Due to \eqref{eq:t=chi}, the anchoring points get mapped to the infinite future and past and are now specified by $\chi$. Hence, in the original geometry, holographic complexity $\mathcal{C}_{\text{dS}}$ is expected to be given by the volume of timelike extremal slices, see Fig. \ref{fig:Ads_vs_dS}. This is the intuition at the basis of the following Section.\\

\noindent \textbf{\emph{Geometric manifestation of microscopic boundary complexity in dS$_2$.--}} In this Section, we deduce the geometric object in the dS limit of sine dilaton gravity whose volume is equal to the Krylov spread complexity in the dual high-energy sector of the DSSYK model. We first discuss this result in detail in the classical limit, and then comment on the validity of the same argument on the quantum level.

Let us start with the complexity duality \eqref{eq:c_duality_general}, valid in general for sine dilaton gravity. By taking the classical limit $q\to1$, we obtain \eqref{eq:c_duality_classicallimit}. Plugging in the dS limit \eqref{limit} into \eqref{eq:c_duality_classicallimit}, the resulting dual gravitational quantity is the classical ERB length $L$ in an effective AdS$_2$ BH geometry with temperature $\beta(\pi-\theta_{\text{dS}})$ with $\theta_{\text{dS}}\ll1$,
\begin{align}\label{classical}
    &\lim_{q\to1} 2\left|\log q\right|C_K(t)_{{\pi-\theta_{\text{dS}}}}=\lim_{q\to1} \mathcal{C}_{\text{eff}}(t)_{{\pi-\theta_{\text{dS}}}}\nonumber\\&=2 \log \left[\cosh \left(t\, \theta_{\text{dS}} /2 \right)\right]-2\log (\theta_{\text{dS}})=L\,.
\end{align}
The temperature $\beta(\pi-\theta_{\text{dS}})=-2\pi/\theta_{\text{dS}}$ in this regime corresponds to the dS inverse temperature $\beta_{\mathrm{dS}}$ \footnote{The interpretation of the formally negative temperature in de Sitter space arises from the fact that $\frac{dS}{dE} < 0 $ so that as the energy increases, the area of the cosmological horizon and hence its entropy decreases. This behavior is a distinctive feature of de Sitter spacetime, where the pure, empty de Sitter vacuum represents a state of maximal entropy. Adding energy reduces the number of accessible microstates of the system. This characteristic behavior is explicitly reflected in the DSSYK spectrum, where the entropy decreases with increasing energy, signaling the instability of the de Sitter saddle point in the DSSYK partition function, as discussed around \eqref{ds_part}.
} and is found by considering the usual limit \eqref{limit} of the sine dilaton one \eqref{dssyk_temp}. From now we will omit the dS subscript on $\theta$. We recall that $L$ is formally obtained by integrating the line element on the spacelike (SL) geodesic between the two asymptotic boundaries of the effective AdS$_2$ geometry,
\begin{align}
\label{eq:L_AdS}
    L=\int_{{\mathrm{SL}\atop |\rho|\,=\,\infty}} \sqrt{ds^2_{\mathrm{AdS}_2}}\,.
\end{align}
The previously discussed map between the effective AdS$_2$ and original dS$_2$ metric via the Weyl rescaling \eqref{weyl} lets us identify the geometric object in the original dS$_2$ metric that has length $L$: the volume of a \emph{timelike} (TL) codimension-1 hypersurface anchored at the \emph{asymptotic future and past} boundaries of dS$_2$ at $\abs{\tau}\to\infty$ \footnote{Both geometric quantities in \eqref{global_transformation} extend over three patches and should be understood as being computed in a maximally extended version of spacetime. Alternatively, as guaranteed by symmetry, they can be understood as twice another quantity that extends from one boundary only to the turning point of the original geodesic, making it amendable to a calculation using a metric that covers two patches, as for example \eqref{dS1}.}:
\begin{equation}
\label{global_transformation}
L\;=\;\underbrace{\int_{{\mathrm{SL}\atop |\rho|\,=\,\infty}} \sqrt{ds^2_{\mathrm{AdS}_2}}}_{=L_{\text{AdS}}:\,{\text{effective geo.}\atop\text{perspective}}}
\;=\;-i \underbrace{\left[\int_{\mathrm{TL}\atop |\tau|\,=\,\infty} \sqrt{ds^2_{\mathrm{dS}_2}}\right]}_{=L_{\text{dS}}:\,{\text{original geo.}\atop\text{perspective}}}.
\end{equation}
While this follows directly from the geometric understanding developed in the last section, an explicit derivation of $L_{\text{dS}}$ starting from the dS$_2$ metric \eqref{dS1} is also provided in the Supplemental Material, and yields
\begin{equation}\label{eq:LendS2}
    L_{\text{dS}}=i\left\{2 \log \left[\cosh \left(\chi\, \theta /2 \right)\right]-2\log (\theta)\right\}\,.
\end{equation}
In the original dS$_2$ geometry, the complexity is a function of the boundary spatial coordinate $\chi$ that specifies the anchoring in the future and past, related to the effective geometry boundary time $t$ via \eqref{eq:t=chi} \footnote{Note that our identification of the DSSYK time with the boundary limit of the Milne spatial coordinate $\chi$, as guided by the sine dilaton duality, is conceptually similar to the identification $\beta \rightarrow -i\ell$ in \cite{Maldacena:2019cbz}, where the dS$_2$ wavefunction can be interpreted as arising from a Lorentzian SYK evolution along the boundary spatial coordinate $\ell$.}. Thereby, we have also confirmed \eqref{global_transformation} explicitly. Taking the square root of the minus sign, which relates AdS$_2$ to dS$_2$, gives rise to the factor of $-i$ in \eqref{global_transformation} multiplying the imaginary length $L_{\text{dS}}$, consistent with the real and positive answer for $L$ \footnote{In taking the square root, we picked the branch with negative sign for coherence of conventions with the higher-dimensional generalization of our proposal, discussed in the next section.}. It is crucial in \eqref{global_transformation} that dS$_2$ is the original geometry of sine dilaton gravity in this limit, because with this information we can \textit{interpret the right hand side of \eqref{global_transformation} as the prescription on how to calculate holographic complexity in dS$_2$}:

\begin{framed}
\noindent 
Krylov spread complexity in the DSSYK model, in the classical limit $q\to1$ and while zooming in on the upper edge of the energy spectrum $\theta\simeq\pi$, is dual to $-i$ times the volume $L_{\text{dS}}$ of the dS$_2$ timelike codimension-1 extremal slice anchored at future and past infinity at spatial coordinate $\chi/2$,
\begin{align}
\label{eq:classical_final_result_two_dim}
    \lim_{q\to1} 2\left|\log q\right|C_K(\chi)_{{\theta\simeq\pi}}=-i\,L_{\text{dS}}(\chi).
\end{align}
We denote this novel timelike notion of cosmological holographic complexity in dS$_2$ by $\mathcal{C}_{\text{dS$_2$}}\equiv -i\,L_{\text{dS}}$.
\end{framed}

To our knowledge, this result is the first observation of an explicit duality between a cosmological holographic complexity and a microscopic boundary complexity.

To conclude this Section, let us now discuss the quantum generalization of \eqref{eq:classical_final_result_two_dim}. This result can naturally be extended to the quantum level (i.e.\ general $q$) thanks to its embedding into the duality between sine dilaton gravity and DSSYK, which holds at the full quantum level for arbitrary values of $\theta$ and $q$. As mentioned in the previous section, keeping $q$ general but taking $\theta\simeq\pi$ gives rise to the regime of the DSSYK model that is understood as dS JT in the quantum regime. The timelike classical length has a natural interpretation as geodesic approximation to the path integral of a matter probe coupled to the gravity theory. As discussed in \cite{Cotler:2019nbi,Maldacena:2019cbz}, one can study gravitational scattering in dS$_2$ by introducing a matter scalar field propagating in global dS$_2$. According to the holographic dictionary for dS spacetime, there exist dual operators located at past and future boundaries. The length proposed in this Letter precisely computes the classical geodesic approximation to the two-point function of these boundary operators, while quantum corrections to its expectation value naturally arise from the gravitational dressing due to interactions with the Schwarzian reparametrization mode emerging on the future and past boundaries of dS$_2$ and have been explicitly computed directly in the dS setting \cite{Cotler:2019nbi,Maldacena:2019cbz}. The quantum analogue of \eqref{eq:classical_final_result_two_dim} can therefore be understood as
\begin{align}
    2\left|\log q\right|C_K(\chi)_{{\theta\simeq\pi}}=-i\langle\hat{L}_{\text{dS}}(\chi)\rangle\,,
\end{align}
where $\langle\hat{L}_{\text{dS}}\rangle$ is the quantum expectation value of the timelike length. For further aspects of the two-dimensional quantum generalization see the Supplemental Material.
\\

\noindent \textbf{\emph{Proposal for cosmological holographic complexity in general dimensions.--}}
The construction of the cosmological holographic complexity $\mathcal{C}_{\text{dS}_2}$ in the previous Section can naturally be extended to higher-dimensional de Sitter spacetimes dS$_{d+1}$ ($d \geq 2$):
\begin{framed}
\noindent In dS$_{d+1}$ ($d\geq 2$), holographic complexity is given by $-i$ times the volume $\mathcal{V}$ of the timelike codimension-1 extremal slice anchored at past and future infinity:
\begin{equation}
\label{eq:higher_d_complexity_proposal}
\mathcal{C}_\text{dS} \equiv \frac{-i\,\mathcal{V}}{G_N \, l_{\mathrm{bulk}}}\ ,
\end{equation}
where $G_N$ the $(d+1)$-dimensional gravitational constant and $l_{\mathrm{bulk}}$ is a length scale for dimensional consistency, taken as the dS radius $l$.

\end{framed}
In this Section, we show that, according to this definition of complexity, higher-dimensional dS spacetimes exhibit the same growth pattern as dS$_2$: a transition from quadratic to linear growth in the spacelike boundary coordinate that parametrizes the slices.

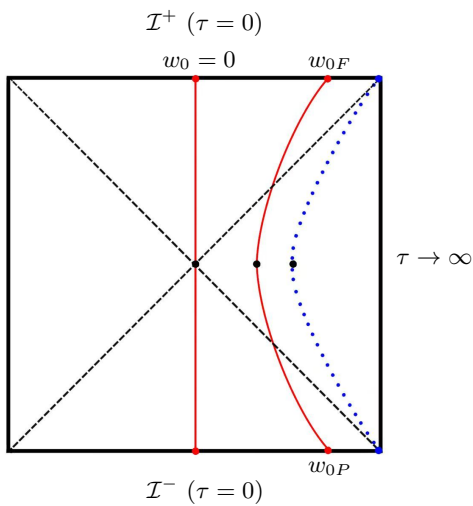
\begin{figure}[h]
  \centering
  \vspace{.5cm}
  \begin{overpic}[width=.8\linewidth]{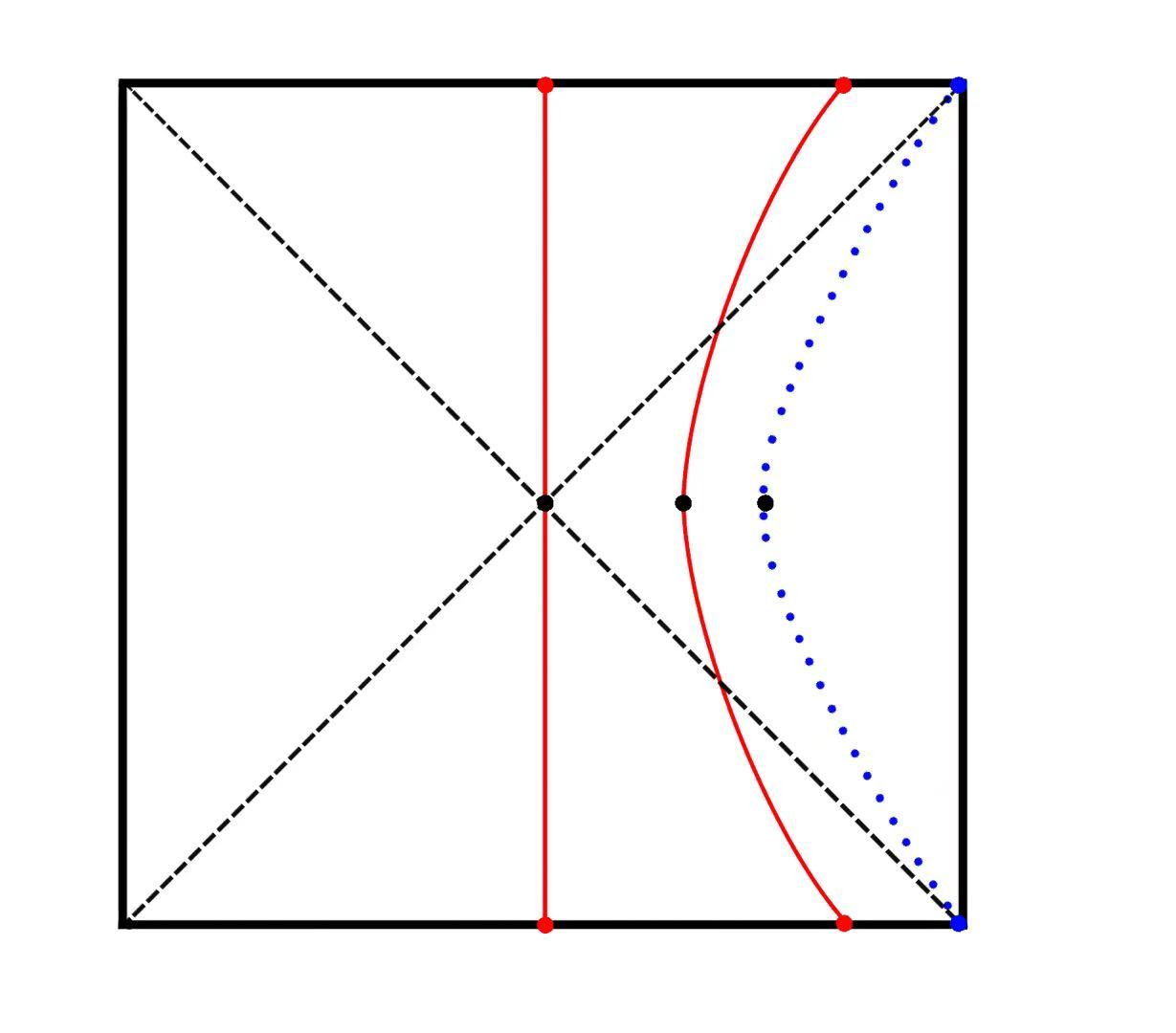}
    \put(37,90){$\mathcal{I}^+$ ($\tau =0$)}
    \put(37,0){$\mathcal{I}^-$ ($\tau =0$)}
    \put(85,45){$\tau \to \infty$}
    \put(40,83){$w_{0}=0$}
    \put(68,83){$w_{0F}$}
    \put(68,5){$w_{0P}$}
  \end{overpic}
  \caption{Penrose diagram of de Sitter spacetime. Future and past infinities $\mathcal{I}^\pm$ are located at $\tau=0$. The cosmological horizons at $\tau=1$ are indicated by black dashed lines. The red curves represent timelike extremal surfaces symetrically anchored at the past infinity point $w_{0P}=w_{0}/2$ and the future infinity point $w_{0F}=w_{0}/2$. Starting from the initial slice $w_0=0$ and increasing $w_0$, the slices approach the accumulation surface, denoted by a blue dotted line. Turning points $(\tau_*,w_*)$ of the extremal surfaces are marked by black dots.}
    \label{fig:ds_penrose} 
\end{figure}

A convenient coordinate choice for the dS$_{d+1}$ metric is
\begin{equation}\label{metric4d}
ds^2=\frac{l^2}{\tau^2}\left(-\frac{d\tau^2}{1-\tau^2}+(1-\tau^2)\,dw^2+d\Omega_{d-1}^2\right) ,
\end{equation}
where $l$ denotes the dS radius and $d\Omega_{d-1}^2$ the line element of the transverse $(d-1)$-sphere. This coordinate system is related to the static patch coordinates for dS$_{d+1}$ by $\tau = l/r$ and $w = t_s/l$, and covers two of the four de Sitter patches \cite{Narayan:2020nsc}. Because of symmetry, this is sufficient to calculate our proposed complexities, which extend over three dS patches, see Fig.~\ref{fig:ds_penrose}.
The range $0 \leq \tau \leq 1$ corresponds to the future (or past) universe, while $\tau > 1$ describes the regions containing the north or south pole, where $w$ reduces to the usual static patch time $t_s$. The hypersurface $\tau = 1$ marks the cosmological horizon, and the asymptotic boundaries at $\mathcal{I}^\pm$ are located at $\tau = 0$.

\begin{figure}
    \centering
    \includegraphics[width=\linewidth]{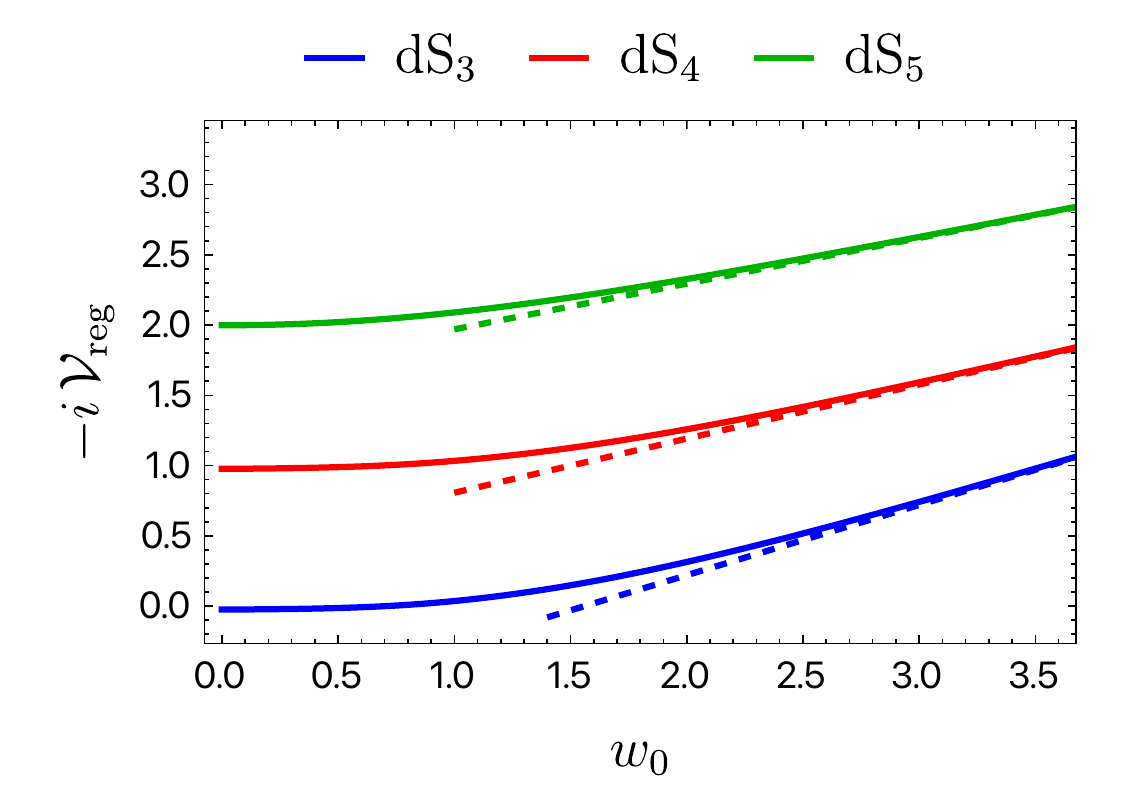}
    \caption{The volume of the timelike extremal slices in dS$_{d+1}$ exhibits quadratic to linear growth as a function of the space coordinate $w_0$ on the Euclidean CFTs at the boundaries. The dotted lines represent the large-$w_0$ growth rate \eqref{growthrate}. The volume of the hypersurfaces has been regularized by subtracting the divergent contribution at $w_0=0$. Moreover, a vertical shift proportional to $d-2$ has been introduced for readability: $\mathcal{V}_\mathrm{reg}=\mathcal{V}(w_0)-\mathcal{V}(0)+i(d-2)$.}
    \label{fig:higher_d_numerics} 
\end{figure}

We will denote the spacelike boundary coordinate as $w(0)\equiv w_0/2$ and anchor the extremal surfaces at symmetric boundary points $w_{0F} = w_{0P}=w_0/2$, as illustrated in Fig.~\ref{fig:ds_penrose}. By using the parametrization $w=w(\tau)$, we can write the volume of the timelike slices as \cite{Narayan:2020nsc}
\begin{equation}
    \mathcal{V} = l^d\Omega_{d-1}\int d\tau\ \frac{i}{\tau^d}\sqrt{\frac{1}{1-\tau^2} - (1-\tau^2)\,\dot{w}^2(\tau)}\,.
\end{equation}
If $\tau_*$ denotes the turning point of the extremal surfaces, defined by $|\dot{w}(\tau_*)|\to\infty$, the growth rate of the volumes $\mathcal{V}$ with the boundary coordinate $w_0$ can be expressed as
\begin{equation}\label{growthrate}
\frac{d\mathcal{V}}{dw_0}=il^d\Omega_{d-1}\,\tau_*^{-d}\sqrt{\tau_*^2-1}+\mathcal{O}(\epsilon).
\end{equation}
See the Supplemental Material for the computation. In the limit $w_0 \to \infty$, \eqref{growthrate} approaches a constant value, as $\tau_*\to \sqrt{d/(d-1)}$. This is understood geometrically from the fact that the slices align more and more with a limiting extremal surface, see Fig.~\ref{fig:ds_penrose}, whose existence was recently highlighted in the context of complexity in dS by \cite{Mohan:2025aiw}. As a consequence, the cosmological volume complexity we have proposed grows linearly at late times. A numerical evaluation of $\mathcal{V}(w_0)$ indeed exhibits quadratic to linear growth and matches with the analytic prediction for the large-$w_0$ growth rate \eqref{growthrate}, see Fig.~\ref{fig:higher_d_numerics}. Moreover, again following~\cite{Mohan:2025aiw}, the growth rate \eqref{growthrate} at large $w_0$ is proportional to temperature times entropy of dS, see the Supplemental Material for details.
\\

\noindent \textbf{\emph{Comparison to earlier work.--}} Our proposal differs significantly from past proposals for holographic complexity in de Sitter spacetimes in the choice of anchoring and the associated causal nature of the extremal hypersurfaces. In AdS/CFT–inspired constructions of dS holographic complexity, spacelike extremal surfaces are typically anchored either at the cosmological horizon~\cite{Susskind:2021esx} or on a stretched horizon just inside it~\cite{Jorstad:2022mls}. In such setups, the complexity typically exhibits hyperfast growth \cite{Susskind:2021esx} (see however~\cite{Aguilar-Gutierrez:2023zqm}).
In contrast, our construction employs timelike hypersurfaces anchored at future and past infinity that evolve with a spacelike coordinate, and the resulting complexity displays linear growth as the spatial boundary coordinate tends to infinity.

A notion of timelike extremal surface as candidate holographic complexity in dS was discussed very recently in \cite{Mohan:2025aiw}, where the same qualitative behavior was found for timelike slices fully contained within the static patch and anchored either at a timelike observer at a dS pole, or at the cosmological horizon. Our proposal is geometrically closely related to the latter case, as the slices we consider can be obtained by extending the horizon-anchored ones of \cite{Mohan:2025aiw} to $\mathcal{I}^\pm$. Because the limiting extremal slice is the same in the two prescriptions, the asymptotic linear growth rate is too: see the Supplemental Material for a more quantitative comparison. The key conceptual difference between the two proposals lies in our choice of anchoring the slices at $\mathcal{I}^\pm$, which entails considering complexity as a function of a spacelike coordinate, and not of time. Ultimately, the main motivation for our proposal is the precise microscopic description in two-dimensional dilaton gravity in terms of Krylov spread complexity, as established in the previous Section.\\

\noindent \textbf{\emph{Interpretation.--}} In the context of dS/CFT \cite{Strominger:2001pn,Balasubramanian:2001nb,Anninos:2011ui,Doi:2022iyj,Kawamoto:2023nki}, the dual to dS spacetime is proposed to be two euclidean CFTs living at the spacelike boundaries located at past and future infinity $\mathcal{I}^\pm$. As a field theoretical quantity, complexity is then expected to depend on the spacelike coordinates of these CFTs. In the dual geometry, these boundary coordinates correspond to the anchoring point of the timelike extremal slices that we introduce as proposal for holographic complexity, see the right picture of Fig.~\ref{fig:Ads_vs_dS}. In that sense, our two-dimensional result for a geometric dual to boundary DSSYK Krylov spread complexity \eqref{eq:classical_final_result_two_dim} and the higher-dimensional extension \eqref{eq:higher_d_complexity_proposal} appear natural from the dS/CFT perspective.

In this interpretation the dual spacetime is probed from $\mathcal{I}^\pm$ and the accessible regions are the past and future Milne patches of Fig.~\ref{fig:Ads_vs_dS}. Bulk reconstruction in dS/CFT has indeed been proposed for these regions \cite{Xiao:2014uea}. The two static patches, instead, are separated from the boundary CFTs by cosmological horizons. As the anchoring coordinate on the boundary grows, the timelike slices that we propose as definition of holographic complexity \emph{measure the increasing size of the region behind the cosmological horizon}, see Fig.~\ref{fig:Ads_vs_dS}. This mechanism is analogous to the AdS/CFT setup depicted on the left of Fig.~\ref{fig:Ads_vs_dS}, where the late-time linear growth of complexity is understood as a measure of the volume beyond the BH horizon \cite{Susskind:2014rva,Gautason:2025ryg}. Moreover, the BH event horizon is conjectured to be the fastest scrambler as it saturates the quantum chaos bound \cite{Maldacena:2015waa}.
A similar growth of complexity in dS
should therefore be connected to the chaotic nature of the cosmological horizon, which has indeed been found by probing dS from a timelike observer in the static patch \cite{Aalsma:2020aib, Giataganas:2021ghs}.

In the outlook we furthermore explore the possibility of an interpretation of our results in terms of a static patch observer.
\\

\noindent \textbf{\emph{Summary and Outlook.--}} 
In this Letter, building on the duality between the high-energy sector of the DSSYK model and the dS regime of sine dilaton gravity, we reported a geometric manifestation of DSSYK Krylov spread complexity as the volume of timelike extremal codimension-1 slices anchored at the infinite future and past in this two-dimensional system. To our knowledge, this constitutes the first observation of an explicit duality of this kind in the context of de Sitter spacetime. Based on this evidence in two dimensions, we proposed holographic complexity to be given by the volume of such extremal timelike surfaces extending from infinite past to future also in higher dimensions. In any dimensions, they exhibit linear growth at large values of the boundary coordinate with a rate proportional to entropy times temperature of dS spacetime. There are several directions for further investigations that this work opens up.

Firstly, there are multiple extensions of the presented calculations that deserve attention. In the two-dimensional case, it would be interesting to study the limit $\theta\simeq\pi/2$, related to two-dimensional flat-space physics \cite{Blommaert:2024ydx}, and identify the geometric flat-space dual to Krylov spread complexity. Moreover, it would be worth considering corrections to the $\Phi\simeq \pi$ regime of sine dilaton gravity: sine dilaton cosmology acts as a regulator of the dS expansion when leaving the strict dS regime \cite{Blommaert:2025rgw}. 
Another interesting aspect, further discussed in the Supplemental Material, is to precisely formulate the quantum statement of the duality established in this Letter. The cosmological quantum corrections given by the Schwarzian mode at future infinity are important for understanding how our proposal embeds into the physics of four-dimensional near-extremal black holes in a dS background. An effective two-dimensional dS JT gravity description, indeed, emerges in the near-horizon, near-extremal limit of the Nariai black hole \cite{Castro:2022cuo}.
Another direction is the study of complexity saturation at late times. Since the $-$AdS contour that sine dilaton naturally suggests around $\Phi=\pi$ is the same gravitational contour that was used in \cite{Cotler:2024xzz,Maldacena:2019cbz} to include higher genus topologies to dS$_2$, it is plausible that the matrix model for dS$_2$ provides the most suitable framework for investigating late-time nonperturbative effects leading to  complexity saturation, in close analogy with the derivation of \cite{Iliesiu:2021ari}. It would also be insightful to explore possible signatures of the switchback effect with our prescription.

Secondly, we provided an interpretation of the spatial boundary coordinate $\chi$ in two dimensions as an effective time coordinate, thanks to the embedding of dS into sine dilaton gravity and DSSYK. This raises the natural question whether it is possible to define a notion of time evolution for complexity directly in dS space, without relying on the effective geometry. The space coordinate at future and past infinity also has the interpretation of a time on the poles of dS, when analytically continuing from Milne to static patch. This location is usually where static patch observers are located \cite{Galante:2023uyf}, motivating the conjectural formula:
\begin{align}\label{conj}
    \lim_{q\to1} 2\left|\log q\right|C_K(\chi)_{{\theta\simeq\pi}}=\mathcal{C}_{\text{dS$_2$}}(\chi)\stackrel{?}{=}\mathcal{C}_{\text{dS$_2$}}(t_{s})\,.
\end{align}
This perspective is consistent with the fact that the generator of spatial translations along the dS boundary becomes the same operator that generates time evolution within the static patch. Consequently, the effective geometry picture provided by sine dilaton gravity may offer a concrete manifestation of the principle that de Sitter dynamics occur in such a way that processes within the static patch are encoded by spatial correlators at infinity \cite{Maldacena:2019cbz}.
If \eqref{conj} was true, the complexity would measure the increasing volume of the portion of the universe that is accessible to the observer at the pole. This would also allow to study possible connections of our investigations to \cite{Aalsma:2020aib, Giataganas:2021ghs}, that discussed the chaotic nature of the dS horizon specifically via a static patch observer. Within the DSSYK framework itself, the backreaction quantum system representing an observer was studied, allowing for a correction to the horizon entropy that reproduces the DSSYK inverse temperature \cite{Tietto:2025oxn}.

Ultimately, the most pressing question is to find microscopic realizations of the proposed holographic complexities in higher dimensions. As much progress has come from the DSSYK model in recent years, it seems especially exciting that a connection of the DSSYK model with dS$_3$ has been proposed in \cite{Narovlansky:2023lfz,Verlinde:2024znh,Tietto:2025oxn}.
We should note here that the dS$_2$ description emerging in the $\Phi \simeq \pi$ regime of the sine dilaton model that we exploited in this paper is not straightforwardly related to the dS$_3$ construction proposed in \cite{Narovlansky:2023lfz,Verlinde:2024znh,Tietto:2025oxn}, where the de Sitter geometry arises at $\theta = \pi/2$. Clarifying this possible connection remains an interesting open problem.\\

\begin{acknowledgments}
We would like to thank Thomas G. Mertens for useful discussions. Furthermore, we specifically thank Vyshnav Mohan and Watse Sybesma for providing and discussing a version of \cite{Mohan:2025aiw} prior to publication which inspired us to think about this subject. T.S. is supported by the Research Foundation - Flanders (FWO) doctoral fellowship 11I5425N. T.S. acknowledges the hospitality of the University of California at Berkeley during the completion of this work. F.O.\ is supported by the Research Foundation Flanders (FWO) doctoral fellowship 1182825N. F.O. thanks for the hospitality the Yukawa Insitute for Theoretical Physics in Kyoto, where this work was completed. JP acknowledges financial support from the European Research Council (grant BHHQG-101040024). Funded by the European Union. Views and opinions expressed are however those of the author(s) only and do not necessarily reflect those of the European Union or the European Research Council. Neither the European Union nor the granting authority can be held responsible for them. M.-T.W.\ acknowledges support from the China Scholarship Council (CSC).

\end{acknowledgments}

\bibliographystyle{bibstyl}
\bibliography{biblio}

\clearpage

\appendix

\clearpage

\onecolumngrid
\section{\large Supplemental material}

\subsection{More on the complexity dictionary in sine dilaton gravity and the DSSYK model}

In this Supplemental Material, we show a short derivation of the key result of \cite{Heller:2024ldz}, namely a precise duality between volume complexity in the effective bulk geometry of sine dilaton gravity and Krylov spread complexity in the DSSYK model. For a complete discussion of this duality, we refer the reader to the original work \cite{Heller:2024ldz} and the references therein.\\

The complexity dictionary is obtained by studying the expectation value of the holographic dictionary \eqref{eq:dictionary_L} for sine dilaton gravity in the time-evolved finite-temperature Hartle-Hawking state
\begin{equation}
\label{eq:state_for_dictionary}
    \ket{\psi(t)_\theta}=Z_{\beta(\theta)}^{-1/2}e^{-i\hat{H}_\text{grav}(t-i\beta(\theta)/2)}\ket{L_\text{eff}=0}\,,
\end{equation}
where $Z_{\beta(\theta)}$ is the thermal partition function. 
The bulk gravitational interpretation of $\bra{\psi(t)_\theta}\hat{L}_\text{eff} \ket{\psi(t)_\theta}$ is the quantum generalization 
of the spacelike wormhole length. To elaborate further, it can be shown that this expectation value can be obtained from the derivative of the sine dilaton gravity two point function of a massive probe with conformal dimension $\Delta$ evaluated at $\Delta=0$, which was introduced in \cite{Iliesiu:2021ari} as notion of length in the quantum regime.

Because geodesic length in two dimensions coincides with the volume of the extremal codimension-1 boundary-anchored slice, the bulk gravitational interpretation of this expectation value is volume complexity in the AdS$_2$ BH effective geometry at time $t$ and inverse temperature $\beta(\theta)$, 
\begin{align}\label{app:CVsine_dilaton}
   \langle\hat L_\mathrm{eff}\rangle=\bra{\psi(t)_\theta}\hat{L}_\text{eff} \ket{\psi(t)_\theta}=\mathcal{C}_\text{eff}(t)_\theta\,.
\end{align}
On the other hand, now taking a boundary viewpoint, the same expectation value \eqref{app:CVsine_dilaton} can also be rewritten as
\begin{align}\label{app:CK_DSSYK}
     L_\mathrm{eff}\rangle=\bra{\psi(t)_\theta}\hat{L}_\text{eff} \ket{\psi(t)_\theta} =2\left|\log q\right|\sum_{n=0}^{\infty} n\left| \langle n |Z_{\beta(\theta)}^{-1/2} e^{-i\hat{T}_\text{DSSYK}(t - i \beta(\theta)/2)}  | 0 \rangle \right|^2\,.
\end{align}
This is precisely a Krylov spread complexity in the boundary DSSYK theory under combined Euclidean and Lorentzian transfer matrix evolution starting from the zero chord state. It is important to highlight that \eqref{app:CK_DSSYK} only can be interpreted as a Krylov spread complexity because \cite{Lin:2022rbf,Rabinovici:2023yex} established that, for evolution with $\hat{H}=\hat{T}_{\text{DSSYK}}$ starting from the zero chord state as reference state $\ket{R}=\ket{0}$, the Krylov basis coincides with the chord basis of DSSYK $\ket{K_n}=\ket{n}$. From the equality of \eqref{app:CVsine_dilaton} and \eqref{app:CK_DSSYK}, we find the complexity dictionary \eqref{eq:c_duality_general} stated in the main text. As said in the beginning, for all further details we refer to \cite{Heller:2024ldz}.

\subsection{Timelike boundary-to-boundary slices in dS$_2$}

\noindent
In this appendix we show that codimension-one timelike extremal slices (i.e.\ timelike geodesics) anchored at future and past infinity of dS$_2$ exhibit the behavior \eqref{eq:LendS2}, hence providing a crosscheck of the argument \eqref{global_transformation}, which is instead obtained from analysis of the role of the Weyl transformation that maps the original to the effective geometry in the dS limit of sine dilaton gravity.

Let us begin with the dS$_2$ metric in Schwarzschild coordinates given by \eqref{dS1}, with $\Phi=r$. It is convenient to use the same coordinates introduced in~\eqref{metric4d}, which span one static and one Milne patch and are related to the Schwarzschild coordinates by $\tau=\theta/r$ and $w=t\,\theta$. The metric in these coordinates is
\begin{equation}
ds^2=-\frac{1}{\tau^2}\left(\frac{d\tau^2}{1-\tau^2}-(1-\tau^2)\ dw^2\right).
\end{equation}
where we see explicitly that $l=1$ in the sine dilaton conventions. The length functional is then given by:
\begin{equation}\label{app:lenfundS2}
L=i\int d\tau\ \frac{1}{\tau} \sqrt{\frac{1}{1-\tau^2}-(1-\tau^2)\ \dot w(\tau)}\,,
\end{equation}
where the quantity under square root is always positive, as the geodesics are timelike. We can introduce a conserved momentum associated to translation invariance along the $w$ direction:
\begin{equation}\label{app:wdotdS2}
E\equiv \frac{\partial\mathcal{L}}{\partial \dot w(\tau)}=\frac{\left(1-\tau ^2\right)\, \dot w(\tau )}{\sqrt{\left(1-\tau ^2\right)^{-1}-\left(1-\tau ^2\right)\, \dot w(\tau )^2}}\qquad\longrightarrow\qquad  \dot w(\tau)=\pm\frac{E\tau}{(1-\tau^2)\sqrt{1+(E^2-1)\tau^2}}\,.
\end{equation}
The turning point of the extremal slices $\tau=\tau_*$ is defined by the condition $|\dot w(\tau_*)|\to \infty$, i.e.
\begin{equation}
\tau_*=\frac{1}{\sqrt{1-E^2}}\ ,\qquad 0\leq E<1\,.
\end{equation}
Using the condition \eqref{app:wdotdS2}, the length is computed from \eqref{app:lenfundS2} by integration from the boundary, regulated as $\tau=\tilde{\epsilon}=\theta\,\epsilon\to 0$, to the turning point $\tau=\tau_*$. By symmetry, the length of the full geodesic is twice this integral:
\begin{equation}\label{app:LendS2}
L=2i\int_{\tilde{\epsilon}}^{\tau_*}d\tau \frac{1}{\tau  \sqrt{1+\left(E^2-1\right) \tau ^2}}=i \left(\log \frac{4}{1-E^2}-2 \log\tilde{\epsilon} \right)+\mathcal{O}(\tilde{\epsilon})\,.
\end{equation}
The boundary coordinate $w(0)\equiv w_0/2$ in terms of which we want to express $L$ is given by
\begin{equation}
w(\tau)=\frac{w_0}{2}+\int_0^\tau\tau'\ \dot w(\tau')\,.
\end{equation}
Since by symmetry $w(\tau_*)=0$, the above expression for $w_0$ evaluated at the turning point gives
\begin{equation}\label{app:boundaryw0}
w_0=-2\int_0^{\tau_*}d\tau'\ \dot w(\tau')=2\,\mathrm{arctanh}\,E\,,
\end{equation}
where we exploit \eqref{app:wdotdS2} again. Note that we are using the convention that $w\to +\infty$ at the right corner of $I^+$, hence here $\dot w<0$ and $w_0>0$. Therefore, $E=\tanh\, (w_0/2)$ and from \eqref{app:LendS2} follows
\begin{equation}\label{app:LendS2bis}
L=2i\log{\cosh{\frac{w_0}{2}}}+2i\log{\frac{2}{\tilde{\epsilon}}}\,.
\end{equation}
Notice that for $\Phi>\theta$ (i.e. $\tau<1$), $\Phi$ parametrizes the time coordinate in the Milne patch, while $t$ parametrizes the spatial Milne coordinate. To be consistent with the standard boundary conditions and holographic renormalization for dS$_2$, the boundary must be located at $\Phi=1/\epsilon \gg1$ in the Schwarzschild coordinates \eqref{dS1} (this is parallel to the boundary conditions in AdS$_2$ \cite{Harlow:2018tqv,Rabinovici:2023yex}).
Since $\tau=\theta/\Phi$, we have chosen the cutoff in $\tau$ to be $\tilde{\epsilon}=\theta \epsilon$, which precisely parametrizes a cutoff at $\Phi=1/\epsilon$. With that, as a consequence of standard holographic renormalization, the non-divergent part $L_\mathrm{dS}\equiv L+2i\log{(\epsilon/2)}$ of the length \eqref{app:LendS2bis} can be written as
\begin{equation}\label{app:lends2}
L_\mathrm{dS}=2i\left( \log \cosh \frac{w_0}{2} -\log \theta \right).
\end{equation}
If we now want to express this result in terms of Milne patch coordinates, we can use that $\omega=t \theta$ and therefore $\omega_0=\chi \theta$ since, in the range $\tau<1$, $t$ parametrizes the Milne spatial coordinate and its boundary limit is precisely $\chi$, in the notation of the main text. In this way, we recover exactly formula \eqref{eq:LendS2} of the main text.

\subsection{Boundary-to-boundary propagator in dS$_2$}

\noindent
In this section, we sketch a semiclassical derivation of the length $L_{\mathrm{dS}}$ of the timelike extremal surface discussed in the main text, proposed there as a measure of de Sitter complexity. This quantity will be obtained via the geodesic approximation to the boundary to boundary propagator of a massive scalar field in the bulk.
To this end, consider a massive scalar field $\psi$ of mass $m$ coupled to dS$_2$ in Poincaré coordinates:
\begin{equation}\label{poin}
ds_{\text{Poincaré}}^{2} = \frac{-d\eta^{2} + dx^{2}}{\eta^{2}} ,
\end{equation}
where $\eta$ denotes the Poincaré time coordinate.
Solving the Klein--Gordon equation for $\psi$ in the background \eqref{poin}, one finds that the matter wavefunctional admits a semiclassical approximation given by a (holographically renormalized) boundary term \cite{Maldacena:2019cbz}:
\begin{equation}\label{mat}
\Psi_{\mathrm{mat}}[\psi_0] = e^{i S_{\psi}} 
\simeq 
\exp\!\left(-i^{2\Delta} \!\int du \int du' 
\frac{\psi_0(u)\psi_0(u')}{|u - u'|^{2\Delta}}\right) ,
\end{equation}
where the scaling dimension $\Delta$ is related to the bulk mass by $m^2 = \Delta(\Delta - 1)$.
Here $\psi_0(u) = \psi(x/\eta_0)$ denotes the fixed boundary value of the bulk field at small cutoff $\eta_0 \ll 1$, with the future boundary located at $\eta_0 = 0$. The wave equation is solved by imposing that the field decays at early times, after an appropriate contour deformation corresponding to the Bunch--Davies (Euclidean) vacuum for the matter field \cite{Maldacena:2019cbz}.  
Equation \eqref{mat} expresses the standard dS/CFT correspondence statement that boundary values of the bulk field at future infinity act as sources for future-boundary operators $\mathcal{O}_{+}(u)$ \cite{Strominger:2001pn}.

However, taking functional derivatives of \eqref{mat} with respect to the boundary sources $\psi_0(u)$ does not yet yield the full physical correlator, since asymptotic diffeomorphisms act nontrivially on the boundary degrees of freedom. The asymptotic dynamics of two-dimensional de Sitter gravity are governed by the Schwarzian action:
\begin{equation}
S_{\mathrm{Sch}}[\varphi] = \int du \, \{ \varphi(u), u \} ,
\end{equation}
where $\{ \varphi(u), u \}$ denotes the Schwarzian derivative and $\varphi(u)$ represents the boundary reparametrization mode at future (and past) infinity \cite{Cotler:2019nbi,Maldacena:2019cbz}. 
The full gravitationally dressed two-point function of boundary operators in this theory is then given by the Schwarzian path integral (modded out by the $SL(2)$ isometry of dS$_2$):
\begin{equation}\label{eq:full_correlator}
\langle \mathcal{O}_{+}(u)\, \mathcal{O}_{+}(u') \rangle 
= 
\int \frac{\mathcal{D}\varphi}{SL(2,\mathbb{R})}
\exp\!\left(i S_{\mathrm{Sch}}[\varphi]\right)
\left(
\frac{\varphi'(u)\,\varphi'(u')}{\bigl[\varphi(u) - \varphi(u')\bigr]^2}
\right)^{\!\Delta} (-i)^{2\Delta} .
\end{equation}
Here, the factor involving $\varphi(u)$ represents the conformal transformation of the matter propagator, while the Schwarzian path integral encodes the contribution of boundary graviton dynamics. Semiclassically, the Schwarzian path integral can be evaluated by the saddle-point approximation, corresponding to the classical solution of the Schwarzian theory. The stationary configuration satisfies $\{\varphi(u),u\}'=0$ and is given by
\begin{equation}\label{saddle}
\varphi(u)=\frac{\beta}{\pi}\tanh\!\left(\frac{\pi u}{\beta}\right) .
\end{equation}
Gravitationally, this saddle admits a natural interpretation: it describes the boundary relation between the Poincaré and Milne spatial coordinates in two-dimensional de Sitter space. This can be seen by rewriting the Poincaré patch \eqref{poin} in lightcone coordinates $u,v = x \pm \eta$, and then performing the change of variables $u=\frac{\beta}{\pi}\tanh\!\left(\frac{\pi \tilde{u}}{\beta}\right), 
v=\frac{\beta}{\pi}\tanh\!\left(\frac{\pi \tilde{v}}{\beta}\right)$.
Under this transformation, the Poincaré metric
\begin{equation}\label{poi2}
ds_{\text{Poincaré}}^{2} = 4\,\frac{du\,dv}{(u-v)^2}\rightarrow ds_{\text{Milne}}^{2} = 
\left(\frac{2\pi}{\beta}\right)^{2}
\frac{-d\hat{t}^{2} + d\chi^{2}}
{\sinh^{2}\!\left(\frac{2\pi \hat{t}}{\beta}\right)} 
\end{equation}
is mapped to the Milne metric, after identifying $\tilde{u}=\hat{t}+\chi$ and $\tilde{v}=\chi-\hat{t}$.  
A further change of variables, defined by
\[
\sinh \tau = \frac{1}{\sinh\!\left(\frac{2\pi \hat{t}}{\beta}\right)}, 
\qquad 
\beta = -\frac{2\pi}{\theta} ,
\]
brings the Milne patch into the form of Eq.~\eqref{eq:milne} used in the main text.

Finally, we can relate the Poincaré spatial coordinate $x$ to the Milne spatial coordinate $\chi$ in the near-boundary limit $\hat{t} = \epsilon \ll 1$ (corresponding to $\tau \gg 1$). This yields
\begin{equation}\label{saddle2}
x=\frac{1}{2}\!\left[
\frac{\beta}{\pi}\tanh\!\left(\frac{\pi(\chi+\epsilon)}{\beta}\right)
+\frac{\beta}{\pi}\tanh\!\left(\frac{\pi(\chi-\epsilon)}{\beta}\right)
\right]
=\frac{\beta}{\pi}\tanh\!\left(\frac{\pi\chi}{\beta}\right)
+\mathcal{O}(\epsilon) ,
\end{equation}
which indeed reproduces the classical saddle \eqref{saddle}. 
This line of reasoning closely parallels the analysis of coordinate frames in AdS JT gravity \cite{Mertens:2022irh}.

Evaluating the Schwarzian-dressed two-point function \eqref{eq:full_correlator} at the classical saddle \eqref{saddle} (or equivalently \eqref{saddle2}), the semiclassical correlator of two operators at future infinity is
\begin{equation}
\langle \mathcal{O}_{+}(u_1)\, \mathcal{O}_{+}(u_2) \rangle 
\simeq 
\left(\frac{-i \theta}{\sinh\left(\frac{\theta\, \delta u}{2}\right)}\right)^{2\Delta},
\end{equation}
where $\delta u = u_1 - u_2$ and the factor of $-i$ arises from the original $-i^{2\Delta}$ in the semiclassical wavefunctional.

To relate the future-future correlator to a future-past correlator, we use an analytic continuation in the complexified spatial coordinate that moves the second operator from future infinity ($\mathcal{I}^+$) to past infinity ($\mathcal{I}^-$):
$u_2 \longrightarrow u_2 + i \frac{\beta}{2} = u_2 - i \frac{\pi}{\theta}$ with $\beta = -2\pi/\theta$. This continuation yields the future-past correlator
\begin{equation}
\label{appeq:twopt_future_past}
\langle \mathcal{O}_{+}(u_1)\, \mathcal{O}_{-}(u_2) \rangle
\simeq 
\left(\frac{\theta}{\cosh\left(\frac{\theta\, \delta u}{2}\right)}\right)^{2\Delta}.
\end{equation}
This property is sometimes referred as the KMS symmetry of de Sitter. Comparing this expression with the geodesic approximation to the Feynman path integral over all boundary anchored paths, in the regime where $m \simeq \Delta$, one obtains:
\begin{equation}
\label{appeq:twopt_future_past}
\langle \mathcal{O}_{+}(u_1)\, \mathcal{O}_{-}(u_2) \rangle
\simeq \int_{\mathrm{anchored \ paths}} \mathcal{D}x e^{-m \int dx}\simeq e^{-m L_{\mathrm{geo}}(\delta u)}
\end{equation}
Setting $\delta u=\chi$ allows to reproduce the geodesic length \eqref{eq:LendS2}. Exploiting this perspective, quantum corrections to the geodesic length \eqref{appeq:twopt_future_past} can be computed perturbatively by expanding around the Schwarzian saddle point \eqref{saddle} as
\begin{equation}
\varphi(u)=\frac{\beta}{\pi}\tanh\!\left(\frac{\pi (u+\varepsilon(u)}{\beta}\right) .
\end{equation}
and using the known form of the propagator $\langle \varepsilon(u_1) \varepsilon(u_2)\rangle$ of the boundary graviton $\varepsilon$ \cite{Cotler:2019nbi,Maldacena:2019cbz}. 

\subsection{Quantum aspects}

\noindent
In this appendix some aspects of the generalization to the quantum regime of the dS limit of sine dilaton gravity are discussed. 
If one takes the exact sine dilaton/DSSYK disk partition function and expands around $\theta \simeq \pi$, one obtains \cite{Blommaert:2024whf,Okuyama:2025hsd}
\begin{equation}
Z_{\mathrm{dS}}(\beta)\propto \int dk \mu(k),
e^{\tfrac{1}{2}|\log q|^2 \beta k^2},
\end{equation}
where $\mu(k)=k \sinh(2\pi k)$ is the JT spectral density. This expression resembles the AdS JT partition function, except for the opposite sign in the Gaussian exponent, which arises from expanding the energy $E=-\cos\theta/2\abs{\log q}$ around $\pi$. This makes it manifest why de Sitter space appears as an unstable saddle.

The integral can be given meaning by performing a Wick rotation $k=-i \tilde{k}$ \cite{Okuyama:2025hsd}, yielding
\begin{equation}\label{ds_part}
Z_{\mathrm{dS}}(\beta)\propto i\left(\frac{2\pi}{|\log q|\beta}\right)^{3/2}
e^{-\tfrac{2 \pi^2}{|\log q|\beta}}
= Z_{\mathrm{AdS}}(-\beta).
\end{equation}
Interestingly, the first equality reproduces the disk amplitude of de Sitter JT gravity computed in \cite{Cotler:2024xzz}, while the second equality shows that it coincides with the AdS JT partition function after $\beta \to -\beta$. This is consistent with the fact that, as discussed in the main text, $\beta_{\mathrm{dS}}=-2\pi/\theta_{\mathrm{dS}}$. The one-loop factor appearing in \eqref{ds_part} arises from the one-loop determinant correction to the Hartle-Hawking wavefunction, which is due to the Schwarzian reparametrization mode living at future infinity in the Milne patch of dS$_2$ \cite{Cotler:2019nbi}. This mode emerges from the spontaneous and explicit breaking of the asymptotic symmetries of dS$_2$ \cite{Maldacena:2019cbz}.

One can proceed in an analogous way for the matter two point function. The AdS JT limit of the sine dilaton gravity/DSSYK matter two point function is known \cite{Blommaert:2023opb}. Because of the symmetry of the integrand over $\theta_{1,2}\rightarrow \pi-\theta_{1,2}$, except for the Boltzman factors, it is tantalizing to write down the following dS limit of the quantum expectation value of the length:
\begin{equation}\label{dsexact}
\langle \hat{L}_{\mathrm{dS}}\rangle=\int d \mu(k_1)d\mu (k_2) e^{\left(\beta/2+it\right)\frac{k_1^2}{2C}}e^{\left(\beta/2-it\right)\frac{k_2^2}{2C}} \frac{\partial}{\partial \Delta} \left[\frac{\Gamma\left(\Delta \pm i k_1 \pm i k_2\right)}{(2C)^{2\Delta}\Gamma (2\Delta)}\right]_{\Delta=0}
\end{equation}
Inspired by the partition function, we could give a meaning to the above integral by performing a double Wick rotation $k_1=-i \tilde{k}_1$, $k_2=-i \tilde{k}_2$. We should remember that the variable $t$ that appears in the direct limit \eqref{dsexact}, due to duality dictionary between sine dilaton gravity and DSSYK, corresponds to the boundary limit of the Rindler wedge time in the effective geometry in Fig.~\ref{fig:Ads_vs_dS}. To interpret \eqref{dsexact} as a dS quantum length expectation value, we should translate $t$ into the the boundary limit (close to future and past infinity) of the spatial coordinate in the Milne patch of dS$_2$.  Despite its very similar form to the standard JT gravity two-point function, to our knowledge an exact expression, at all orders in the gravitational coupling constant $C=1/\abs{\log q},$ of the dS two point function and thus of the length expectation value of the form \eqref{dsexact} has not appeared in the literature so far, while its AdS JT counterpart has been computed in many ways \cite{Mertens:2022irh}. In \cite{Maldacena:2019cbz} it was left as an open question whether, in the dS case, one can also integrate out the Schwarzian reparametrization modes exactly as in AdS. It would be nevertheless interesting, by using the techniques of \cite{Griguolo:2021zsn}, to check the proposed expression \eqref{dsexact} against the one loop corrections to the matter gravitational scattering problem \cite{Cotler:2019nbi}, given by the interaction of the matter particle with the Schwarzian mode living at future and past infinity.

A related question regarding \eqref{dsexact} is whether one might obtain it directly by performing a canonical quantization of dS in terms of the length $L_{\mathrm{dS}}$ of the extremal surface connecting the the asymptotic boundaries (future and past infinity), i.e. the vertical line in Fig.~\ref{fig:Ads_vs_dS} we have been considering in this paper, and its conjugate momentum $P_{\mathrm{dS}}$. In the spirit of Harlow and Jefferis \cite{Harlow:2018tqv}, one would obtain an Hamiltonian of the form:
\begin{equation}\label{hami}
H_{\mathrm{dS}}=-P_{\mathrm{dS}}^2/2-2 e^{-L_{\mathrm{dS}}},
\end{equation}
which is precisely minus the corresponding Liouville Hamiltonian describing the time evolution of two-sided Cauchy slices in AdS JT gravity \cite{Harlow:2018tqv}. The fact that the de Sitter energy appears with the opposite sign to the corresponding AdS JT ADM energy can be understood from the dimensional reduction of near-extremal black holes. In this context, the two-dimensional energies measure the deviation above extremality, and the requirement of avoiding naked singularities in the four-dimensional geometry fixes the relative signs of the mass deviations, $\delta M$ \cite{Castro:2022cuo}. We notice that this construction would exactly lead to the expression \eqref{dsexact}, since the eigenvectors of \eqref{hami} would be unaltered but the eigenvalues would get a minus sign. The major obstacle in this construction would be that the corresponding hamiltonian flow would be along the spatial direction \cite{Maldacena:2019cbz}. However, the effective geometry, which played a central role in the quantization of sine dilaton gravity, might suggest a natural reinterpretation of this apparent puzzle: the operator generating spatial translations along the boundary in the dS$_2$ system corresponds to a standard Hamiltonian operator generating evolution along the effective time in the effective geometry. 

A final comment on the de Sitter limit of sine dilaton gravity concerns the natural emergence of a Schwarzian mode description at the boundary of dS$_2$, which appears closely tied to the original DSSYK theory from which we started. While it is well known that a Schwarzian theory emerges in the deep IR limit of the SYK model, and consequently of DSSYK, it is not established whether an analogous description arises in the deep UV, which is the regime we are considering in this Letter. Nevertheless, since the SYK spectrum is statistically symmetric under flipping the sign of the energy \cite{Maldacena:2019cbz}, corresponding in our setting to the symmetry $\theta \to \pi-\theta$, we are led to speculate that the very high-energy sector of the SYK spectrum may behave similarly to the low-energy sector. Our results provide evidence in support of this intriguing possibility.

We leave these questions, that go beyond the scope of the current work, for future investigations.

\subsection{Growth rate in dS$_{d+1}$}

\noindent
In this appendix we detail the computation of the growth rate for the volume of codimension-1 timelike extremal slices in dS$_{d+1}$ ($d\geq 2$). As explained in the main text, we write the metric as
\begin{equation}\label{app:metric4d}
ds^2=\frac{l^2}{\tau^2}\left(-\frac{d\tau^2}{1-\tau^2}+(1-\tau^2)\,dw^2+d\Omega_{d-1}^2\right),
\end{equation}
and choose to parametrize the extremal slices as $w=w(\tau)$. The action principle that defines the volume of the slices is then given by:
\begin{equation}
\mathcal{V}=l^d\Omega_{d-1}\int\mathcal{L}\,,
\qquad
\mathcal{L}=\frac{i}{\tau^d}\sqrt{\frac{1}{1-\tau^2}-(1-\tau^2)\,\dot w^2(\tau)}\,,
\end{equation}
with conserved quantity
\begin{equation}\label{app:Edef}
E\equiv \frac{\del \mathcal{L}}{\del \dot w}=-\frac{\left(1-\tau ^2\right)\, \dot w(\tau )}{\tau ^{d}\sqrt{\left(1-\tau ^2\right)^{-1}-\left(1-\tau ^2\right)\, \dot w(\tau )^2}}
\qquad\longrightarrow\qquad
\dot w(\tau) = \pm \frac{E \tau^{d}}{(1-\tau^2)\sqrt{1-\tau^2+E^2\tau^{2d}}}\,.
\end{equation}
Note that \eqref{app:Edef} reduces to \eqref{app:wdotdS2} for $d=1$, es expected. The turning point of the extremal slices $\tau=\tau_*$ is defined by the condition $|\dot w(\tau_*)|\to \infty$, i.e.
\begin{equation}\label{app:turningp}
1-\tau_*^2+E^2\tau_*^{2d}=0\,.
\end{equation}
By symmetry, the volume of extremal slices in this parametrization is given by twice the volume of the half-slice extending from future infinity, which we regulate as $\tau=\epsilon\to 0$, to the turning point $\tau=\tau_*$:
\begin{equation}\label{app:Vdensity}
\mathcal{V}=2i\,l^d \Omega_{d-1}\int_\epsilon^{\tau_*} d\tau\ \frac{\tau^{-d}}{\sqrt{1-\tau^2+E^2\tau^{2d}}}\equiv \int_\epsilon^{\tau_*}d\tau\ f_\mathcal{V}(\tau,E)\,.
\end{equation}
The boundary coordinate $w(0)\equiv w_0/2$ in terms of which we want to express $\mathcal{V}$ is given by
\begin{equation}\label{app:boundaryW}
w_0=-2\int_0^{\tau_*}d\tau\ \dot w(\tau)\,,
\end{equation}
compare with \eqref{app:boundaryw0}.
Note that we are using the convention that $w\to +\infty$ at the right corner of $I^+$, hence here $\dot w<0$ and $w_0>0$. The growth rate can be evaluated by exploiting the chain rule as
\begin{equation}
\frac{d\mathcal{V}}{dw_0}=\frac{d\mathcal{V}}{dE}\frac{dE}{dw_0}\,,
\end{equation}
where, using the notation of \eqref{app:Vdensity},
\begin{equation}\label{app:chainrule1}
\frac{d\mathcal{V}}{dE}=\frac{d}{dE}\int_\epsilon^{\tau_*} d\tau\ f_\mathcal{V}(\tau,E)=f_\mathcal{V}(\tau_*,E)\,\frac{d\tau_*}{dE}+\int_\epsilon^{\tau_*}d\tau\ \frac{d}{dE}f_\mathcal{V}(\tau,E)\,,
\end{equation}
and similarly, from \eqref{app:boundaryW} with \eqref{app:Edef},
\begin{equation}\label{app:chainrule2}
\frac{dE}{dw_0}=\left(\frac{dw_0}{dE}\right)^{-1}=\left[-2\frac{d}{dE}\int_0^{\tau_*} d\tau\ \dot w(\tau,E)\right]^{-1}
=-\frac{1}{2}\left[\dot w(\tau_*,E)\,\frac{d\tau_*}{dE}+\int_0^{\tau_*}d\tau\ \frac{d}{dE}\dot w(\tau,E)\right]^{-1}.
\end{equation}
In both equations above, $E=E(\tau_*)$ is expressed as a function of $\tau_*$ from \eqref{app:turningp}. Note that both terms \eqref{app:chainrule1} and \eqref{app:chainrule2} are inherently divergent or zero at $\tau=\tau_*$, but their product is finite. In practical terms, the growth rate is more easily extracted by expressing \eqref{app:chainrule1} and \eqref{app:chainrule2} as a function of $\tau=\tau_*+\delta$, and then taking the limit $\delta \to 0$. The result, to leading order in the cutoff $\epsilon\to 0$, is indeed given by \eqref{growthrate},
\begin{equation}
\frac{d\mathcal{V}}{dw_0}=il^d\Omega_{d-1}\,\tau_*^{-d}\sqrt{\tau_*^2-1}+\mathcal{O}(\epsilon)\,.
\end{equation}
As a side comment, note that the small-$w_0$ regime exhibits quadratic growth because of time-reflection symmetry with respect to $w_0=0$: the first derivative of the volume $d\mathcal{V}/dw_0=0$, hence when expanded around $w_0=0$, it has an expansion of the form $\mathcal{V}=\mathcal{V}_0+\mathcal{V}_2w^2+\mathcal{O}(w^4)$ \cite{Carmi:2017jqz}.

\subsection{Geometrical analogies with the proposal by Mohan and Sybesma \cite{Mohan:2025aiw}}

\noindent
In this appendix, we compare in more detail the geometrical aspects of our proposal for complexity in dS$_{d+1}$ to a different proposal recently presented in \cite{Mohan:2025aiw}. More in detail, we will show that the two share the same limiting extremal surface, which is reached as the variable the complexity depends on goes to infinity. We will also show, as a consequence, that the two asymptotic growth rates are the same. The important conceptual differences between the two proposals are instead illustrated in the main text of the Letter.

The limiting extremal surface in \cite{Mohan:2025aiw} lies at
\begin{equation}
\lim_{t\to\infty} r_*(t)=l\sqrt{\frac{d-1}{d}}
\end{equation}
where $(t,r)$ are the usual static patch coordinates. This is equivalent to our limiting slice
\begin{equation}
\lim_{w_0\to\infty}\tau_*(w_0)=\sqrt{\frac{d}{d-1}}\,,
\end{equation}
as it can be immediately seen using the coordinate transformation $\tau=l/r$. Hence, the limiting extremal surface is the same, and as a consequence the asymptotic growth rates are dictated by the same geometric property in both proposals. As discussed for example in \cite{Gautason:2025ryg} for the AdS case and commentend on in \cite{Mohan:2025aiw} for the dS case, the asymptotic growth of holographic volume complexity proposals is indeed typically dominated by the contribution beyond the horizon. Indeed, for the growth rate \cite{Mohan:2025aiw} has
\begin{equation}\label{growth_rateW}
\frac{d\mathcal{V}}{d\Delta t}=
\frac{d\mathcal{V}}{d \Delta v}=\Omega_{d-1} r_*^{d-1}\sqrt{1-r_*^2}
\end{equation}
depending whether the slices are anchored at the static patch observer worldline ($\Delta t=t_F-t_P$) or at the cosmological horizon ($\Delta v/2=v_F=-u_P$). The relation $w=t/l$ between the static patch time $t$ and the $w$ coordinate allows to express our complexity as a function of the static patch time $t$, as well. Using the correspondence between $\Delta t$ and the boundary coordinate $w_0$, i.e.\ $w=t/l$, leads to $\Delta t=lw_0$ and
\begin{equation}\label{app:w_vs_t}
\frac{d\mathcal{V}}{d\Delta t}=\frac{1}{l}\frac{d\mathcal{V}}{dw_0}\,,
\end{equation}
which, once plugged in \eqref{growth_rateW} together with the coordinate transformation $r_*=l/\tau_*$, matches \eqref{growthrate}, besides a factor of $i$ that in the conventions of \cite{Mohan:2025aiw} is contained in their imaginary reference scale. In this convention, one could interpret the factor of $-i/l_{\text{bulk}}$ in our definition of cosmological holographic complexity \eqref{eq:higher_d_complexity_proposal} as such, although \emph{ab initio} we propose the prefactor of $-i$ by analogy with the two-dimensional case, where it arises as square root of the Weyl factor $-1$ that relates effective and original geometry in the specific lower-dimensional example \eqref{eq:LendS2}. Moreover, again following~\cite{Mohan:2025aiw}, the growth rate \eqref{growthrate} at large $w_0$ can be rewritten as
\begin{equation}
    \lim_{w_0\to\infty}\frac{d\mathcal{V}}{dw_0}=i\frac{8\pi G_N l^2}{\sqrt{d}}\left(\frac{d-1}{d}\right)^{\frac{d-1}{2}}ST \propto S T \, ,
\end{equation}
where $S$ and $T$ are the dS entropy and temperature,
\begin{equation}
    S = \frac{\Omega_{d-1} l^{d-1}}{4 G_N} \,,
    \qquad 
    T = \frac{1}{2\pi l} \,.
\end{equation}

\begin{figure}[h]
    \centering
    
    \subfigure{\includegraphics[width=.45\linewidth]{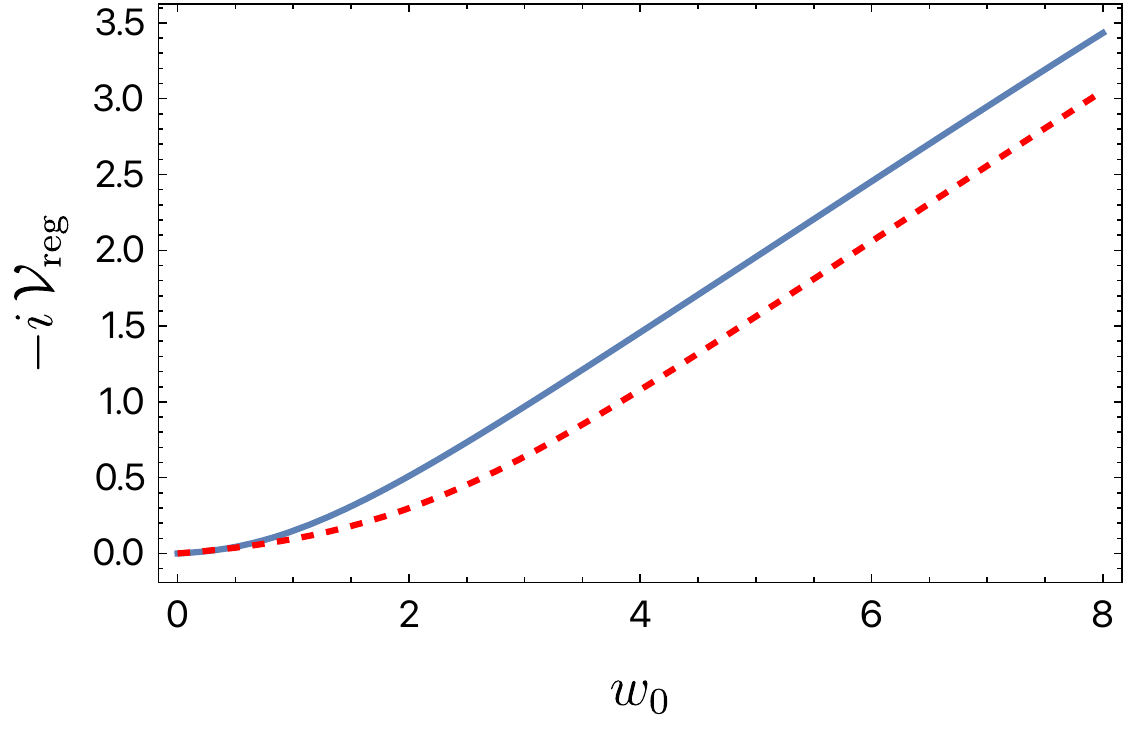}
    }
    \subfigure{
    \hspace{.05\linewidth}
    \begin{overpic}[width=.45\linewidth]{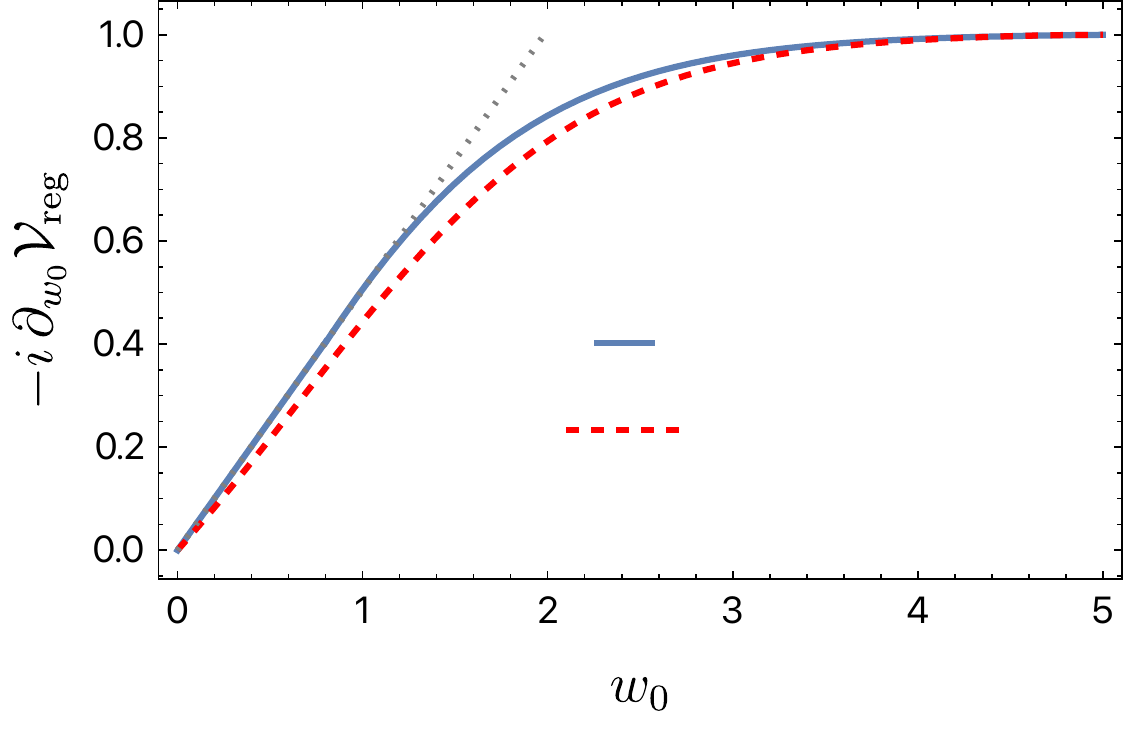}
    \put(65,33){Our proposal}
    \put(65,25.5){Proposal by \cite{Mohan:2025aiw}}
    \end{overpic}
    }
    
    \caption{Comparison between the proposals for complexity in dS$_{d+1}$ introduced in the present Letter and in \cite{Mohan:2025aiw}. In these plots, $d=2$. \textbf{Left panel}: the imaginary part of the regularised volume (compare with Fig.~\ref{fig:higher_d_numerics}) is compared by expressing both quantities as a function of $w_0$ as described in the body of this Supplemental Material. The asymptotic growth rate as $w_0\to \infty$ is the same, even if the small $w_0$ behavior differs. \textbf{Right panel}: the same comparison for the derivative of $\mathcal{V}_\mathrm{vreg}$, showing in both cases a transition from quadratic to linear growth.}
    \label{fig:comparison_proposals}
\end{figure}

Equivalently, we can relate our complexity to the proposal where the slices are anchored at the cosmological horizon by expressing the infalling coordinate
\begin{equation}
v=t+r_*+\frac{i\pi l}{2}
\end{equation}
as a function of $w$ on the cosmological horizon $r=l$ or $\tau=1$. Note that we have to include the imaginary quarter-of-a-period shift, since we are considering $v$ in the region $r>l$ or $0<\tau<1$. In the $(\tau,w)$ coordinates we have
\begin{equation}
v=t+\int \frac{dr}{1-\left(\frac{r}{l}\right)^2}+\frac{i\pi l}{2}=l \left(w+\operatorname{arctanh}\frac{1}{\tau }+\frac{i\pi}{2}\right),
\end{equation}
and if we relate for example $w$ and $w(0)\equiv w_0/2$ along a null radial path, with the care of picking the infalling solution we find
\begin{equation}
\frac{dw}{d\tau}=\pm\frac{1}{1-\tau^2}\qquad\longrightarrow\qquad
w-\frac{w_0}{2}=-\operatorname{arctanh} \tau\,.
\end{equation}
On the cosmological horizon, which we appproach from $\tau\to1^-$, 
\begin{equation}
\Delta v=2v_F=2\lim_{\tau\to1^-}\ l\left(\frac{w_0}{2}-\operatorname{arctanh}\tau+\operatorname{arctanh}\frac{1}{\tau}+\frac{i\pi}{2}\right)= lw_0\,,
\end{equation}
allowing to compare the growth rates in the same way as with the static patch time. See the numerical comparison in the case of cosmological horizon anchoring in Fig.~\ref{fig:comparison_proposals}: note that the two quantities are not identical, but share the same growth rate and overall qualitative quadratic to linear behavior.

\end{document}